\documentclass[10pt,A4]{article}
\usepackage{amssymb}
\usepackage{graphicx}
\usepackage{wrapfig}
\usepackage{natbib}
\begin{document}

\title{Modeling the spectrum of gravitational waves \\
    in the primordial Universe}

\author{M. S. Santos $^1$\thanks{
e-mail: \texttt{msoares@astro.iag.usp.br}} \ , 
        S. V. B. Gon\c{c}alves $^2$\thanks{
e-mail: \texttt{sergio@cce.ufes.br}} \ ,
        J. C. Fabris $^2$\thanks{
e-mail: \texttt{fabris@cce.ufes.br}} \ and \\
        E. M. de Gouveia Dal Pino $^1$\thanks{
e-mail: \texttt{dalpino@astro.iag.usp.br}} \\\\ 
\mbox{\small $^1$ Instituto de Astronomia, Geof\'{\i}sica
e Ci\^encias Atmosf\'ericas, Universidade de S\~ao Paulo} \\ 
\mbox{\small $^2$ Departamento de Física, Universidade Federal do 
Esp\'{\i}rito Santo}}

\maketitle

\begin{abstract}
Recent observations from type Ia Supernovae and  from cosmic
microwave background (CMB) anisotropies have revealed that most of
the matter of the Universe interacts in a repulsive manner,
composing the so-called dark energy constituent of the Universe.
The analysis of cosmic gravitational waves (GW) represents,
besides the CMB temperature and polarization anisotropies, an
additional approach in the determination of parameters that may
constrain the dark energy models and their consistence. In recent
work, a generalized Chaplygin gas model was considered in a flat
universe and the corresponding spectrum of gravitational waves was
obtained. The present work adds a massless gas component to that
model and the new spectrum is compared to the previous one. The
Chaplygin gas is also used to simulate a $\Lambda$-CDM model by
means of a particular combination of parameters so that the
Chaplygin gas and the $\Lambda$-CDM models can be easily
distinguished in the theoretical scenarios here established. The
lack of direct observational data is partialy solved when the
signature of the GW on the CMB spectra is determined.
\vspace{0.7cm}
\end{abstract}

\maketitle


\section{Introduction}

A large number of dark energy candidate models have been proposed
since that the SNeIa \citep{Riess:1998,Permultter:1998,Tonry:2003} and
CMB \citep{Tegmark:2004} experiments  revealed the accelerating
expansion of the universe. In order to explain the observational
data all these models consider an equation of state with negative
pressure. The  {\it cosmological constant} models consider a
simple equation of state, $p=-\rho$, which however results a huge
discrepancy with the  data \citep{Sahni:2004}. Another class of
models considers a scalar field, called {\it quintessence}
\citep{Caldwell/Doran:2004}, a third class assumes a perfect
fluid with a negative pressure which is proportional to the
inverse of the energy density, the {\it Chaplygin gas}
\citep{Gorini:2004},  and finally a forth class, considers  an {\it
X-fluid} with an equation of state $p= - \omega \rho$, where $\omega$
is a positive constant, known as phantom energy when $\omega>1$
\citep{Sahni:2004}.

The distinction among these models must be done by means of
combined observations which  up to the present have been unable to
define what description is more appropriate. Gravitational waves
represent one of the potential tools that may in near future offer
an additional way to constrain the cosmological parameters and
distinguish among the models, since  gravitons decouple in a very
early time in the universe history. In this paper, we study the
spectrum of gravitational waves due to the X-fluid model and
compare the results with a previous work \citep{Fabris:2004}, where
the cosmological constant and the generalized Chaplygin gas models
were taken into account. In section \S\ref{outline of the model}
these models are described; in \S\ref{GW equations} the GW
equations are presented. The spectra analysis is left to
\S\ref{spectra analysis}, which is followed by the conclusions of
this work.

\section{Outline of the model}
\label{outline of the model}
We consider a flat, homogeneous and isotropic universe described
by the Friedman-Robertson-Walker metric, which
makes the Einstein's equations to assume the form
\begin{equation}
\label{einsteineq1}
 \left(\frac{\dot{a}}{a}\right)^2 =
 \frac{8\pi G}{3} \ (\rho_m + \rho_{de}),
\end{equation}
\begin{equation}
\label{einsteineq2}
 \frac{\ddot{a}}{a}+2\left(\frac{\dot{a}}{a}\right)^2=-8\pi G
 \ (p_m + p_{de}),
\end{equation}
where $a$ is the scale factor of the universe, while $\rho_m$ and
$\rho_{de}$ are the pressureless fluid and the dark energy
densities, respectively. The pressures $p_m$ and $p_{de}$ of the
fluids are related with their densities by the  equations of state
$p_m=0$, and $p_x=\omega \rho_x$, with $\omega<0$ (in case of an
X-fluid) or $p_c=-A / \rho_c^\alpha$, with $A,\alpha>0$ (in case
of Chaplygin gas), respectively. We take the scale factor today as
the unity, $a_0=1$ (The subscripts $0$, according to the current
notation, indicate the present values of the quantities.) and
rewrite the dark energy density for both cases as:
\begin{equation} \label{rho_x}
\rho_x=\rho_{x_0} a^{-3(\omega+1)},
\end{equation}
\begin{equation} \label{rho_ch}
 \rho_c=\rho_{c_0}
 \left[\bar{A} + \frac{(1 - \bar{A})}{a^{3(\alpha + 1)}}
 \right]^{\frac{1}{\alpha + 1}} \ , \
 \bar{A}=\frac{A}{\rho_{c_0}^{\alpha + 1}}.
\end{equation}
With equations (\ref{rho_x}) and (\ref{rho_ch}), we write the
Einstein's equations in the form
\begin{equation}
\label{adot x} \frac{\dot{a}}{a}=H_0 \left(
\frac{\Omega_{m_0}}{a^3} + \frac{\Omega_{x_0}}{a^{3(\omega+1)}}
\right)^{1/2}
\end{equation}
\begin{equation}
\label{addot x} \frac{\ddot{a}}{a}=-\frac{1}{2} H_0^2 \left(
\frac{\Omega_{m_0}}{a^3} + \Omega_{x_0}
\frac{1+3\omega}{a^{3(\omega+1)}} \right)^{1/2},
\end{equation}
for the X-fluid case, and
\begin{equation}
\label{adot ch}
 \frac{\dot{a}}{a} = H_0 \left[\frac{\Omega_{m_0}}{a^3} +
 \Omega_{c_0} \left(\bar{A} + \frac{1 - \bar{A}}{a^{3 (\alpha +1)}}
 \right)^{\frac{1}{\alpha+1}} \right]^{1/2},
\end{equation}
\begin{equation}
\label{addot ch}
 \frac{\ddot{a}}{a} = H_0^2 \left[- \frac{\Omega_{m_0}}{2a^3}
 + \Omega_{c_0}  \left(\bar{A} + \frac{1 - \bar{A}}
 {a^{3(\alpha + 1)}} \right)^{\frac{1}{\alpha + 1}}  \left(1 -
 \frac{3(1-\bar{A})}{2a^{3(\alpha + 1)}} \left(\bar{A} +
 \frac{1 - \bar{A}}
 {a^{3(\alpha + 1)}} \right)\right) \right],
\end{equation}
for the Chaplygin gas. The Hubble constant $H_0$ is defined as
$H_0=\dot{a_0}/a$ and, once we are restricted to a flat universe,
the fractions of pressureless matter and dark energy gas today,
$\Omega_{m_0}$ and $\Omega_{de_0}$, respectively, obey the
relation $\Omega_{m_0} + \Omega_{de_0} = 1$.

With these two last equations we are able to write the GW
amplitude differential equation as a function of the observable
parameters $H_0$, $a$, $\Omega_{m_0}$ and $\Omega_{de_0}$, and of
the dark energy fluid parameters: $\bar{A}$, $\alpha$ or $\omega$.
It is also important to remark that \\
(i) \ \ \ if $\bar{A}=0$, then the Chaplygin gas behaves like the
pressureless fluid and the situation is the same as if we had set
$\Omega_{m_0}=1$; \\
(ii) \ \ on the other hand, for $\bar{A}=1$ it behaves like the
Cosmological Constant  and therefore, we can simulate the
$\Lambda$-CDM scenario; \\
(iii) \ the X-fluid is equal to the Cosmological Constant for
$\omega=-1$.

Among the many possible cases produced by the combinations of the
parameters we have chosen a few important ones which are indicated
in table 1. These choices  will be sufficient for the aim of the
present analysis, where we compare the spectra due to each model.

\begin{center}
\begin{table}
\centering
\begin{tabular}{cccccc}
\hline 
& $\omega$ & $\bar{A}$ & $\alpha$ & $\Omega_{de_0}$ &
  $\Omega_{m_0}$  \\[2pt] \hline 
Pressureless Fluid & -- & -- & -- & 0 & 1 \\[2pt] \hline
Cosmological Constant & -1 & 1 & -- & 0.96 & 0.04 \\[2pt] 
&  &  &  & 0.7 & 0.3 \\[2pt] \hline
Generalized Chaplygin Gas & -- & 0.5 & 1 & 0.96 & 0.04 \\[2pt] 
&  &  &  & 0.7 & 0.3 \\[2pt] 
&  &  & 0.5 & 0.96 & 0.04 \\[2pt] 
&  &  &  & 0.7 & 0.3 \\[2pt] 
&  &  & 0 & 0.96 & 0.04 \\[2pt] 
&  &  &  & 0.7 & 0.3 \\[2pt] \hline
X-fluid & -0.5 & -- & -- & 0.7 & 0.3  \\[2pt] \hline
X-fluid (Phantom Energy)& -10 & -- & -- & 0.7 & 0.3  \\[2pt]
\hline
\end{tabular}
\caption{Models chosen for the analysis.}
\end{table}
\end{center}

\section{GW equations}
\label{GW equations} Cosmological gravitational waves are obtained
through little perturbations $h_{ij}$ on the metric. Hence, the
tensor $g^{(0)}_{ij}$, related to the unperturbed metric, is
replaced by $g_{ij} = g^{(0)}_{ij} + h_{ij}$ and the resulting
expression is \citep{JW:W72} :
\begin{equation}
\label{gw1}
 \ddot{h} - \frac{\dot{a}}{a} \dot{h}
 + \left( \frac{k^2}{a^2} - 2 \frac{\ddot{a}}{a} \right) h = 0 ,
\end{equation}
where $k$ is the wave number times the velocity of light ($k=2 \pi
c/\lambda$), the dots indicate time derivatives and $h(t)$ is
defined as: $h_{ij}(\vec{x},t)=h(t)Q_{ij}$, where $Q_{ij}$
are the eigenmodes of the Lagrangian operator, such that
$Q_{ii}=\partial_k Q_{ki}=0$.

Proceeding with a variable transformation from time to the scale
factor $a$, and representing the derivatives with respect to $a$
by primes, equation (\ref{gw1}) assumes the form
\begin{equation}
\label{gw2}
 h'' + \left( \frac{\ddot{a}}{\dot{a}^2} - \frac{1}{a} \right)h'
 + \frac{1}{\ddot{a}^2} \left( \frac{k^2}{a^2}
 - 2 \frac{\ddot{a}}{a} \right)h = 0
\end{equation}
and, by the application of the background equations (\ref{adot
ch}) and (\ref{addot ch}), or (\ref{adot x}) and (\ref{addot x}),
which concern to the dark energy fluid, into (\ref{gw2}), one can
easily express $h$ in terms of the parameters of the model.

\subsection{Generalized Chaplygin gas}

Let us perform the last operation mentioned above and, in order to
find $h$ as a function of the redshift $z$, let us (i) use the
known relations $1+z=\frac{a_0}{a}$, $a_0=1$; (ii) perform a
second
 change of variables (from $a$ to $z$); and (iii) take back the dots
to indicate, from now on, the new integration variable. These
operations result in
\begin{equation}
\label{gw_DE}
 \ddot{h} + \left[\frac{2}{1+z}+\frac{3}{2}(1+z)^2 \frac{f_1}{f_2}
 \right]\dot{h} + \left[\frac{k^2}{f_2}-\frac{2}{(1+z)^2}+3(1+z)
 \frac{f_1}{f_2} \right]h = 0 \ ,
\end{equation}
\begin{equation}
\label{f1}
 f_1=\Omega_{m_0}+\Omega_{c_0}(1-\bar{A})(1+z)^{3\alpha}
 \left[\bar{A}+(1-\bar{A})(1+z)^{3(\alpha+1)} \right]^
 {\frac{1}{\alpha+1}-1} \ ,
\end{equation}
\begin{equation}
\label{f2}
 f_2=\Omega_{m_0}(1+z)^3 + \Omega_{c_0} \left[\bar{A}+
 (1-\bar{A})(1+z)^{3(\alpha+1)} \right]^{\frac{1}{\alpha+1}},
\end{equation}
where k is redefined to absorb the Hubble constant, i.e., $k =
2\pi c / H_0  \lambda$ and, therefore, $h$ is a dimensionless
quantitiy.

\subsection{X-fluid}
To obtain the equivalent equations for the X-fluid model, we use
the same procedure described above. The result may be written in
the same form as in (\ref{gw_DE}), with $f_1$ and $f_2$ redefined
as:
\begin{equation}
\label{f1x}
 f_1=\Omega_{m_0}+\Omega_{x_0}(1+\omega)(1+z)^{3\omega} \ ,
\end{equation}
\begin{equation}
\label{f2x}
 f_2=\Omega_{m_0}(1+z)^3 + \Omega_{x_0}(1+z)^{3(\omega+1)} \ .
\end{equation}
One can easily notice that, if $\omega=-1$ and $\bar{A}=1$, both
cases converge to the same equation and then reproduce  the
$\Lambda$-CDM scenario.

\section{GW spectra analysis}
\label{spectra analysis}
The power spectrum of gravitational waves, defined as
\begin{equation}
\label{powerspectrum}
 \frac{d\Omega_{GW}}{d \ln \nu}=|h_0(\nu)|\nu^{5/2}
\end{equation}
[where $h_0(\nu)=h(0)$ and $\nu=H_0 k / 2\pi$], is generally
obtained from the solutions of (\ref{gw_DE}) which are found by
means of a numerical algorithm specifically created for this
problem. In the particular case of the $\Lambda$-CDM model, an
analytical result is also possible  and this fact is used to
verify the accuracy of the calculation.

For each of the cases of interest, we have assigned some common
parameters, namely the initial conditions $h(z_i)=\nu^{-3}
10^{-5}$, $\dot{h}(z_i)=\nu^{-2} 10^{-5}$, $z_i=4000$; the range of
frequencies $10^{-18} Hz \leq \nu \leq 10^{-15} Hz$, which is a
small part of the range of observational interest for GW 
that goes up to up to $10^{10} Hz$; and the
normalization constant imposed by the CMB \citep{Riazuelo/Uzan:2000}
\begin{equation}
\label{constr}
 \frac{d\Omega_{GW}}{d \ln \nu}\Big|_{10^{-18}} \leq 10^{-10} .
\end{equation}

The resulting spectra are presented in figures (1-11) and we can
see that they all have a similar behavior: a strong oscillatory
shape; a peak at $\nu \approx 0$ (due to the initial condition,
which is proportional to $k^{-3}$); after a very rapid decrease,
the spectra increase slowly. Besides these features, it is
important to remark that: \\\vspace{5mm}
\begin{center}
  \begin{minipage}[t]{0.47\linewidth}
    \includegraphics[width=\linewidth]{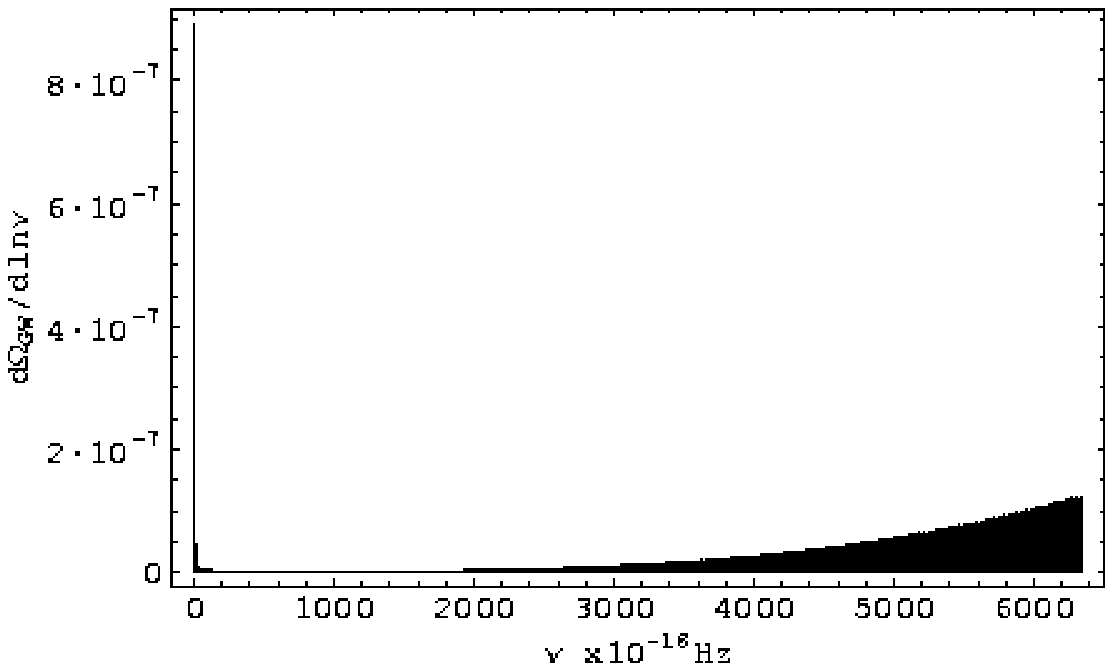}
    {\footnotesize {\bf Figure 1.} GW spectrum for
       $\Omega_{c_0}=0$ and
       $\Omega_{m_0}=1$.}
  \end{minipage} \hfill
\end{center}
\begin{tabular}{c c}
  \begin{minipage}[t]{0.47\linewidth}
    \includegraphics[width=\linewidth]{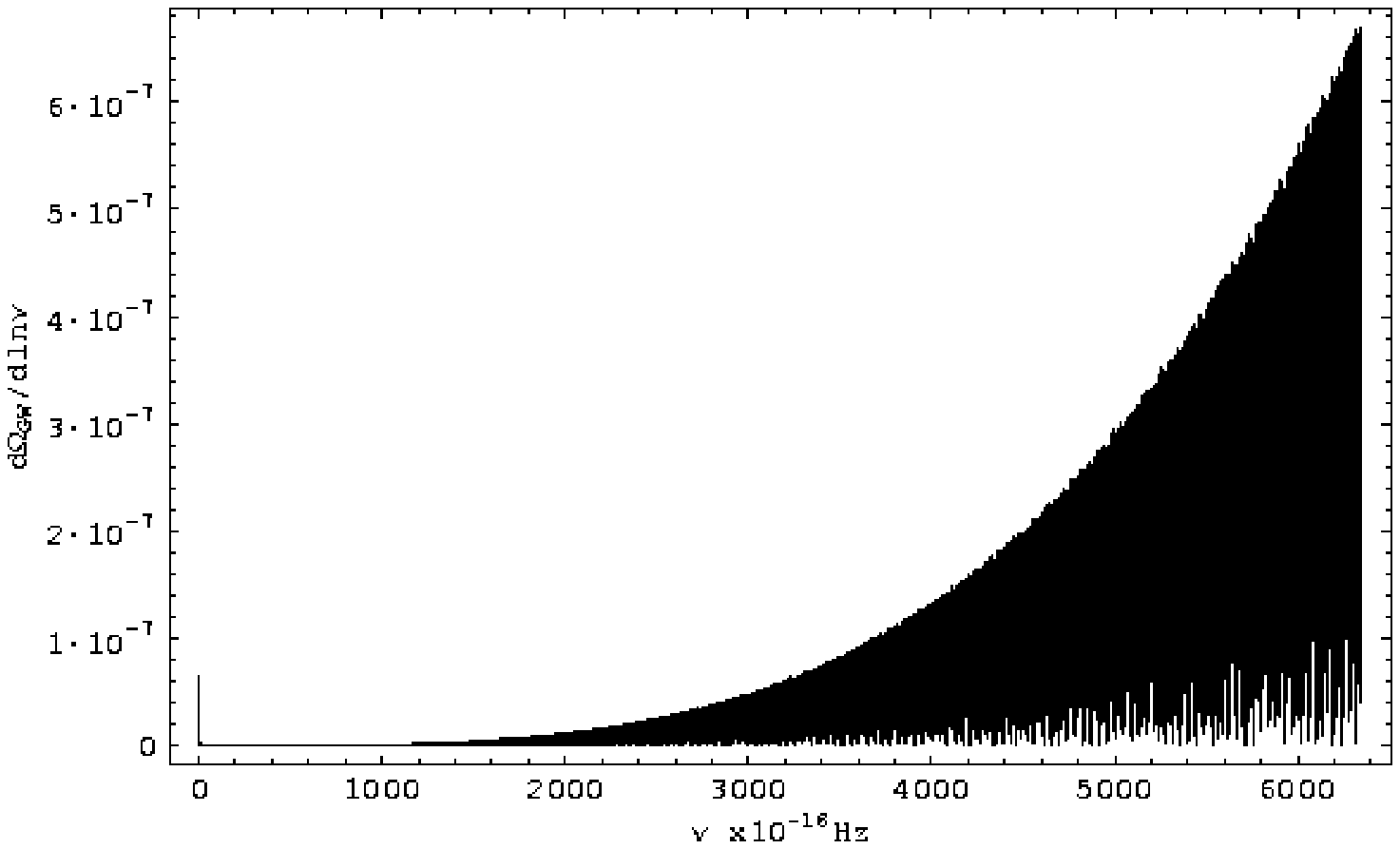}
    {\footnotesize  {\bf Figure 2.} GW spectrum for
       $\bar{A}=1$,
       $\Omega_{c_0}=0.96$ and
       $\Omega_{m_0}=0.04$.}
  \end{minipage} \hfill &
  \begin{minipage}[t]{0.47\linewidth}
    \includegraphics[width=\linewidth]{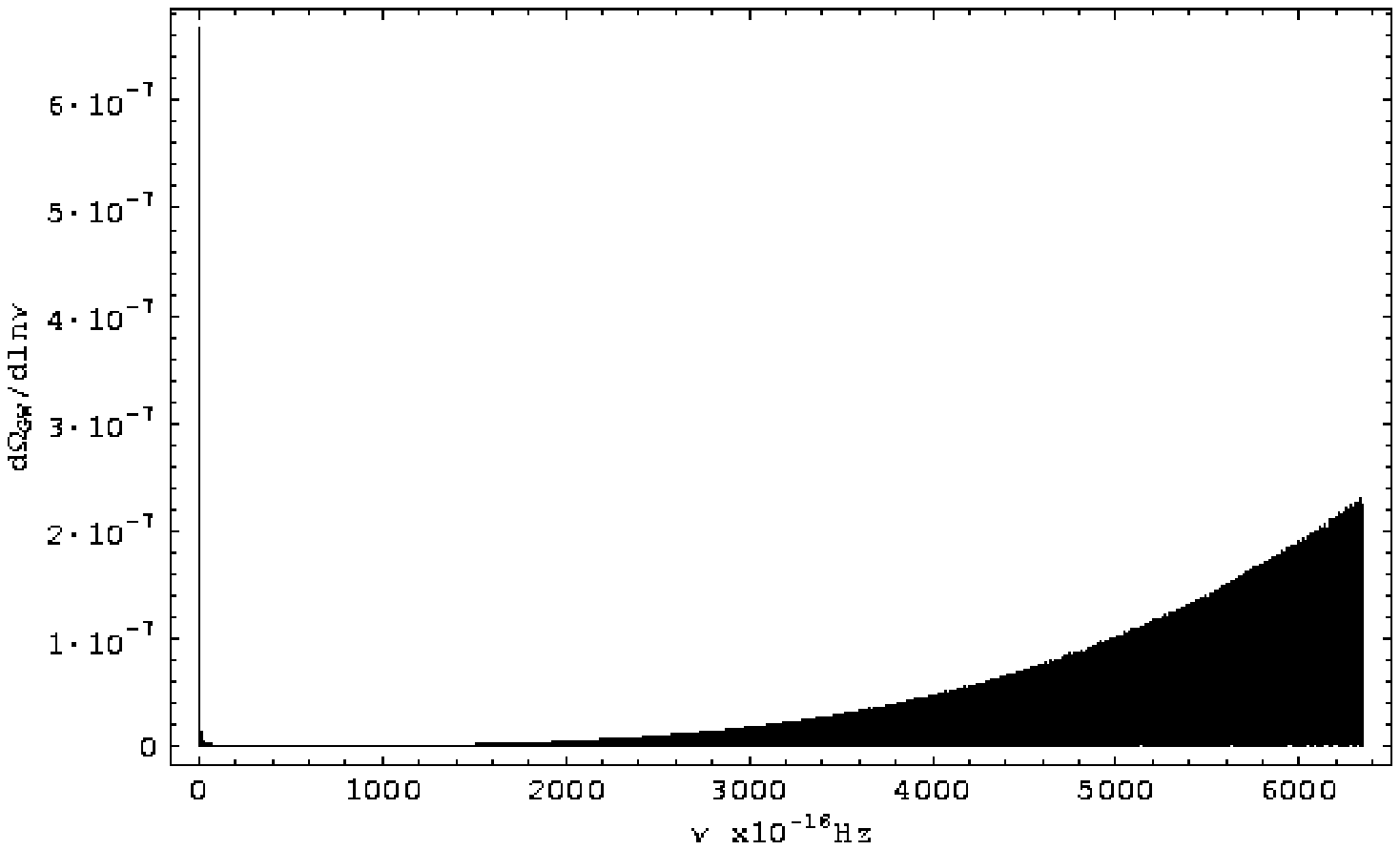}
    {\footnotesize  {\bf Figure 3.} GW spectrum for
       $\bar{A}=1$,
       $\Omega_{c_0}=0.7$ and
       $\Omega_{m_0}=0.3$.}
  \end{minipage} \hfill
\end{tabular} \\\vspace{5mm}

(i) \ \ The plot (Fig. 1) is for a pressureless pure model and we
can see that the spectrum corresponding to a Cosmological Constant
dominated universe (Fig. 2) grows faster with the frequency
reaching a value three times greater. Still referring to a
$\Lambda$-CDM model, figure (Fig. 3) shows a behavior similar to
the pressureless fluid, but the growing amplitude is just a little
larger.
\\\vspace{5mm}

\begin{tabular}{c c}
  \begin{minipage}[b]{0.47\linewidth}
    \includegraphics[width=\linewidth]{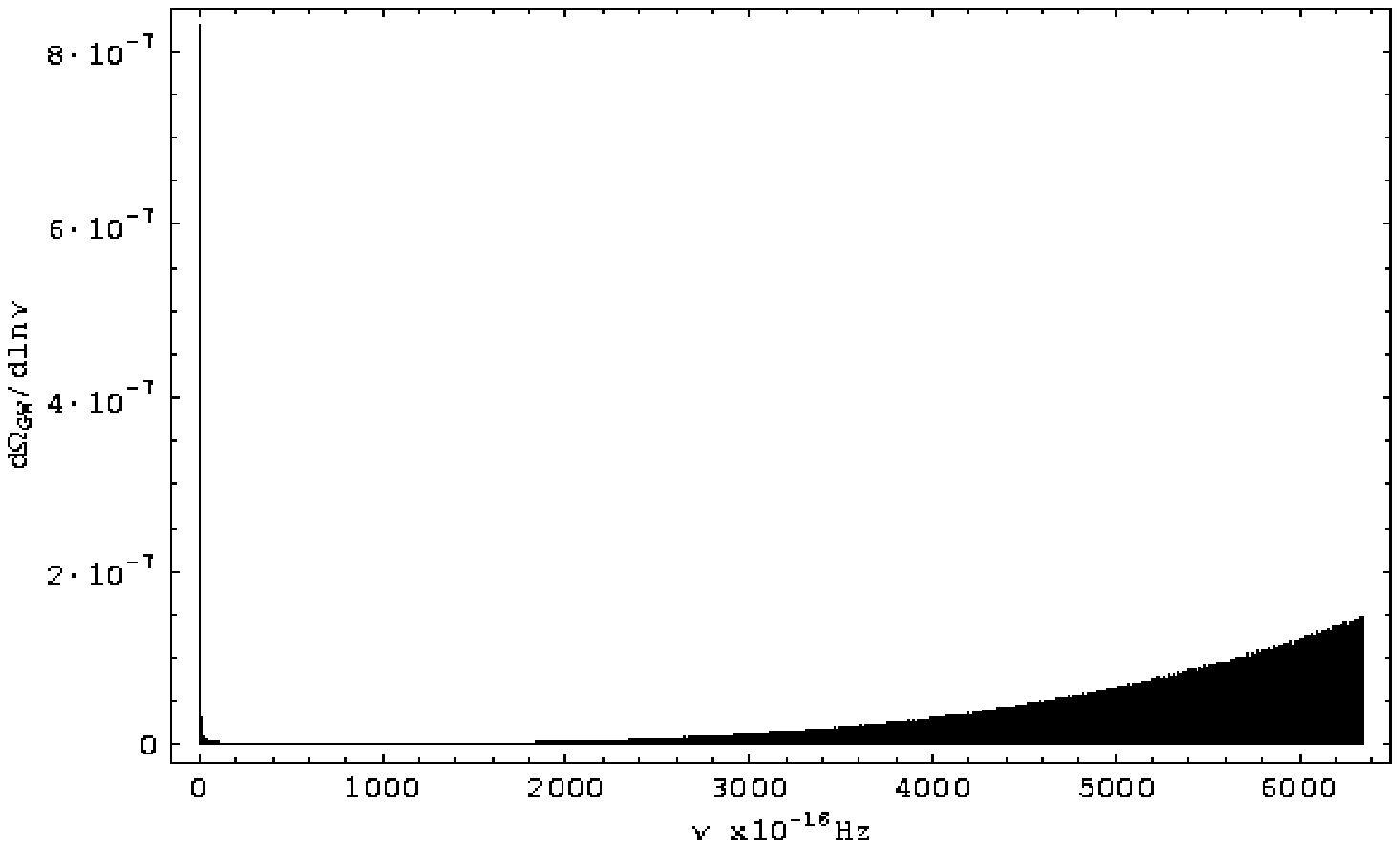}
    {\footnotesize {\bf Figure 4.} GW spectrum for
       $\bar{A}=0.5$,
       $\alpha=1$,
       $\Omega_{c_0}=0.96$ and
       $\Omega_{m_0}=0.04$.}
  \end{minipage} \hfill &
  \begin{minipage}[b]{0.47\linewidth}
    \includegraphics[width=\linewidth]{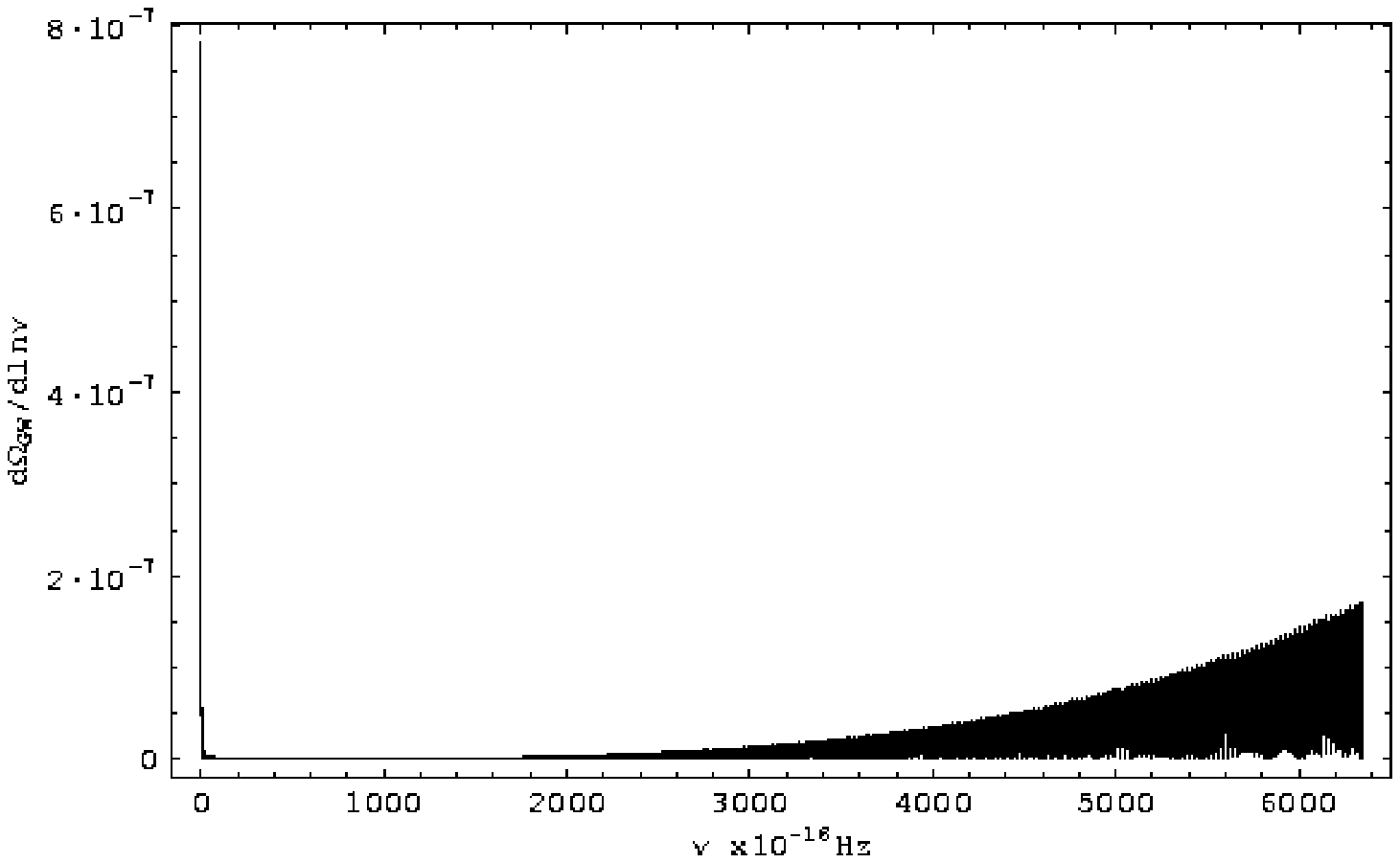}
    {\footnotesize {\bf Figure 5.} GW spectrum for
       $\bar{A}=0.5$,
       $\alpha=1$,
       $\Omega_{c_0}=0.7$ and
       $\Omega_{m_0}=0.3$.}
  \end{minipage} \hfill
\end{tabular} \\\vspace{5mm}

\begin{tabular}{c c}
  \begin{minipage}[b]{0.47\linewidth}
    \includegraphics[width=\linewidth]{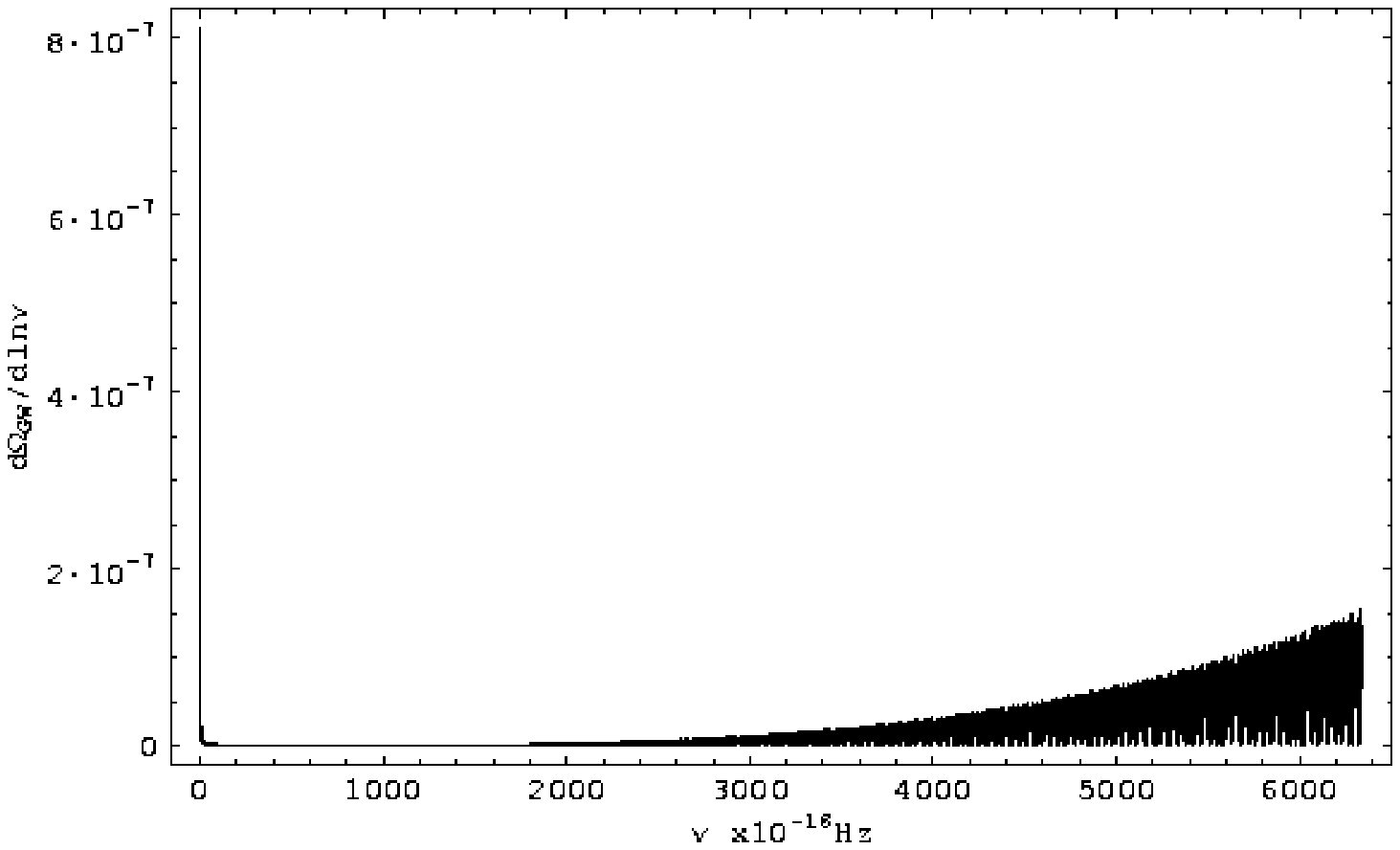}
    {\footnotesize {\bf Figure 6.} GW spectrum for
       $\bar{A}=0.5$,
       $\alpha=0.5$,
       $\Omega_{c_0}=0.96$ and
       $\Omega_{m_0}=0.04$.}
  \end{minipage} \hfill &
  \begin{minipage}[b]{0.47\linewidth}
    \includegraphics[width=\linewidth]{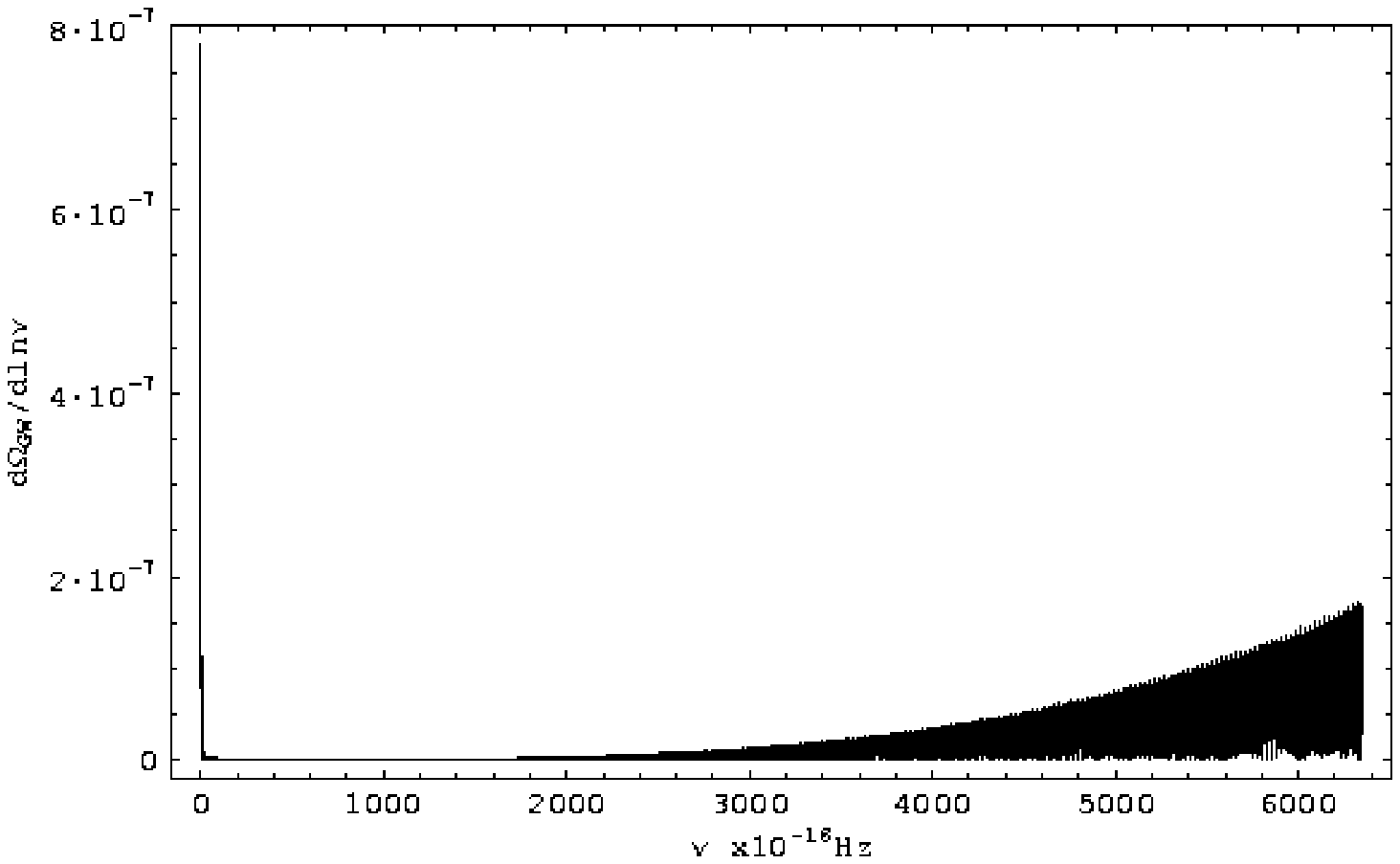}
    {\footnotesize {\bf Figure 7.} GW spectrum for
       $\bar{A}=0.5$,
       $\alpha=0.5$,
       $\Omega_{c_0}=0.7$ and
       $\Omega_{m_0}=0.3$.}
  \end{minipage} \hfill
\end{tabular} \\\vspace{5mm}

\begin{tabular}{c c}
  \begin{minipage}[b]{0.47\linewidth}
    \includegraphics[width=\linewidth]{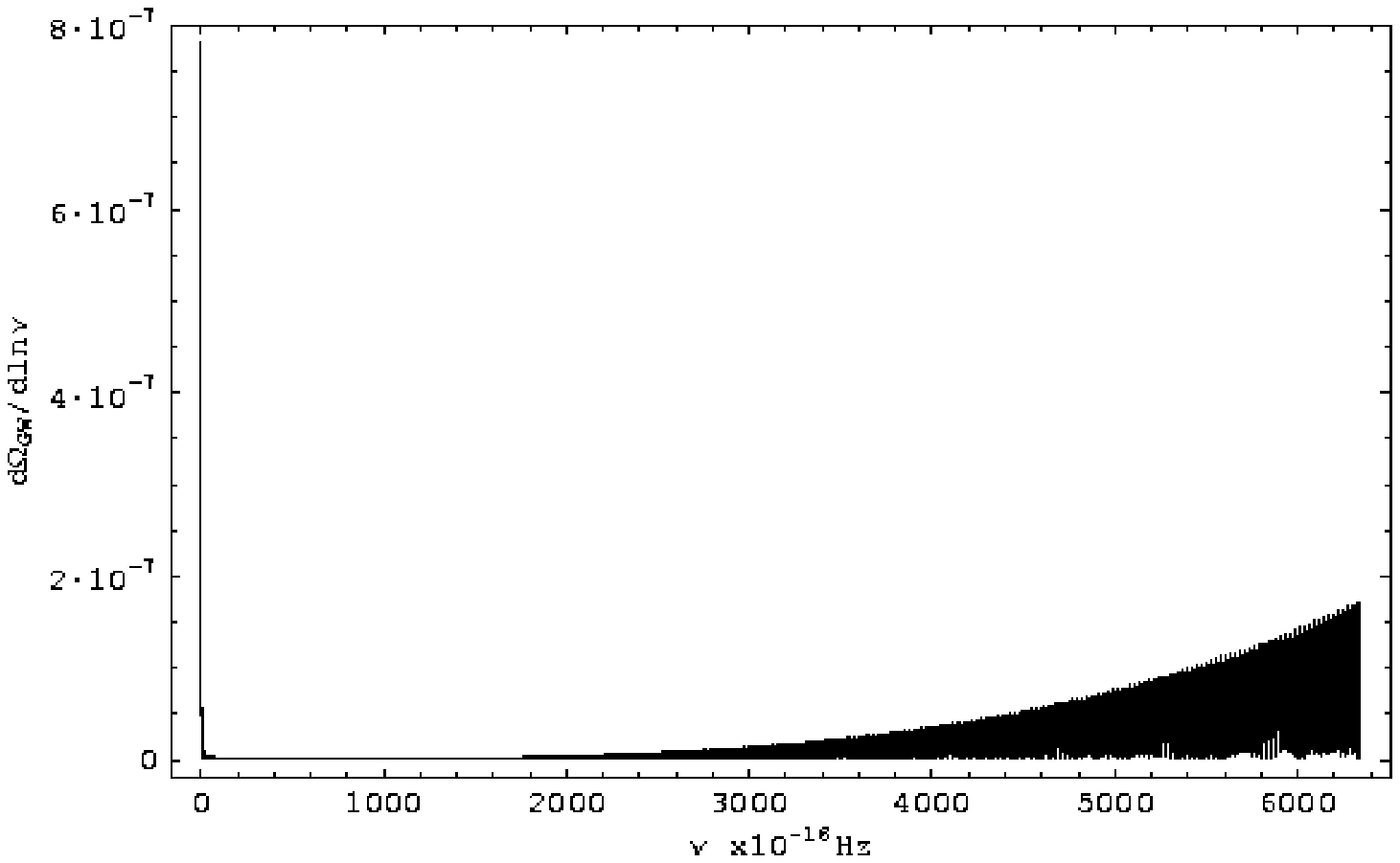}
    {\footnotesize {\bf Figure 8.} GW spectrum for
       $\bar{A}=0.5$,
       $\alpha=0$,
       $\Omega_{c_0}=0.96$ and
       $\Omega_{m_0}=0.04$.}
  \end{minipage} \hfill &
  \begin{minipage}[b]{0.47\linewidth}
    \includegraphics[width=\linewidth]{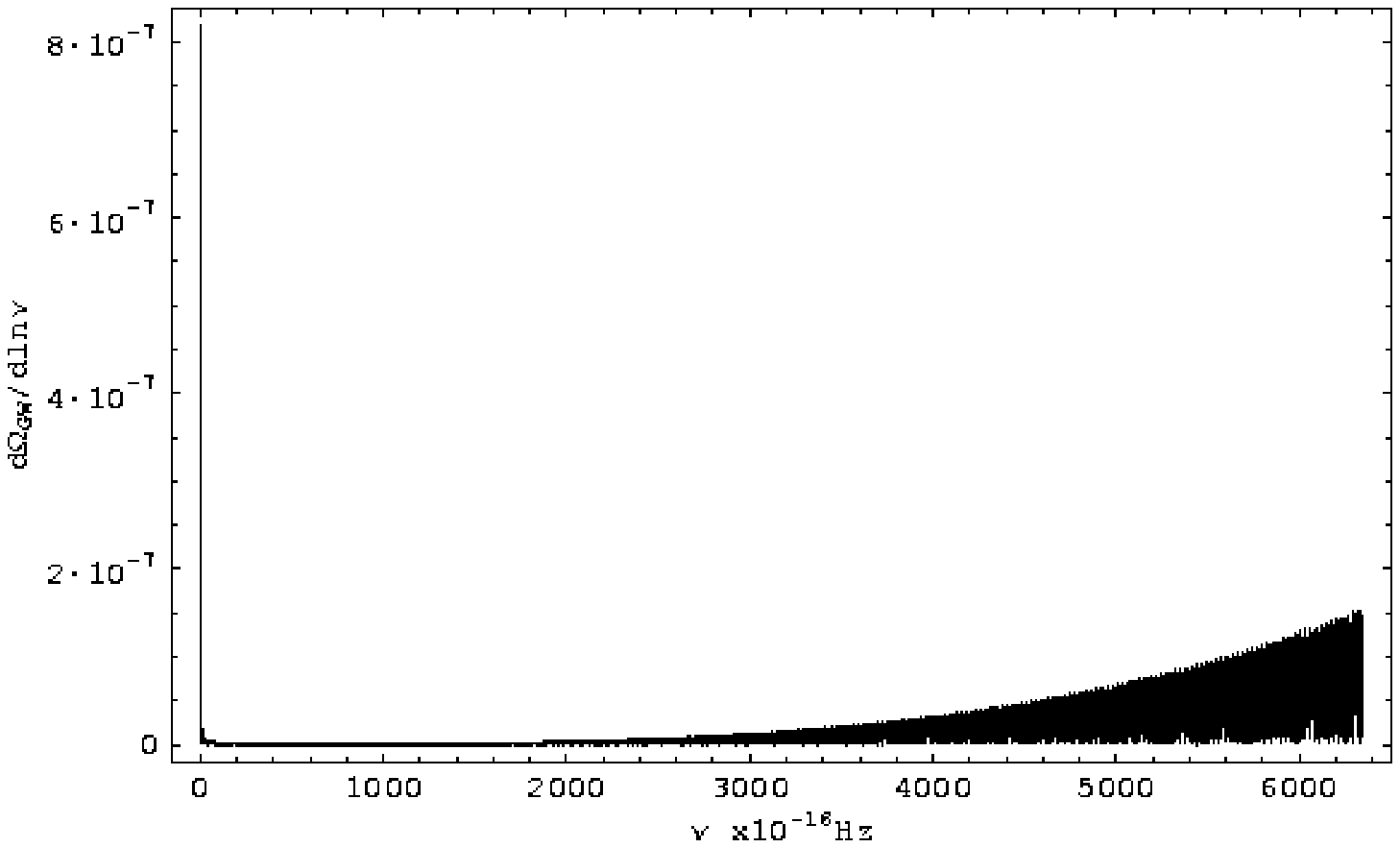}
    {\footnotesize {\bf Figure 9.} GW spectrum for
       $\bar{A}=0.5$,
       $\alpha=0$,
       $\Omega_{c_0}=0.7$ and
       $\Omega_{m_0}=0.3$.}
\end{minipage} \hfill
\end{tabular} \\\vspace{5mm}

(ii) \ The graphics (4-9), however,
are not so easily distinguishable. They seem to have no dependence
on the $\Omega_{c_0}$ parameter and this may be interpreted as a
consequence of the Chaplygin gas property of interpolation between
dark-matter and dark-energy. 

(iii) Figures (4) and (5) refer to the
Chaplygin gas in its traditional form -- with the equation of
state $p=-A/\rho$, since $\alpha=1$ -- while
(6-9) correspond to the generalized Chapligyn gas and the six
plots together show that the spectrum is not affected
by $\alpha$. \\\vspace{5mm}

\begin{tabular}{c c}
  \begin{minipage}[b]{0.47\linewidth}
    \includegraphics[width=\linewidth]{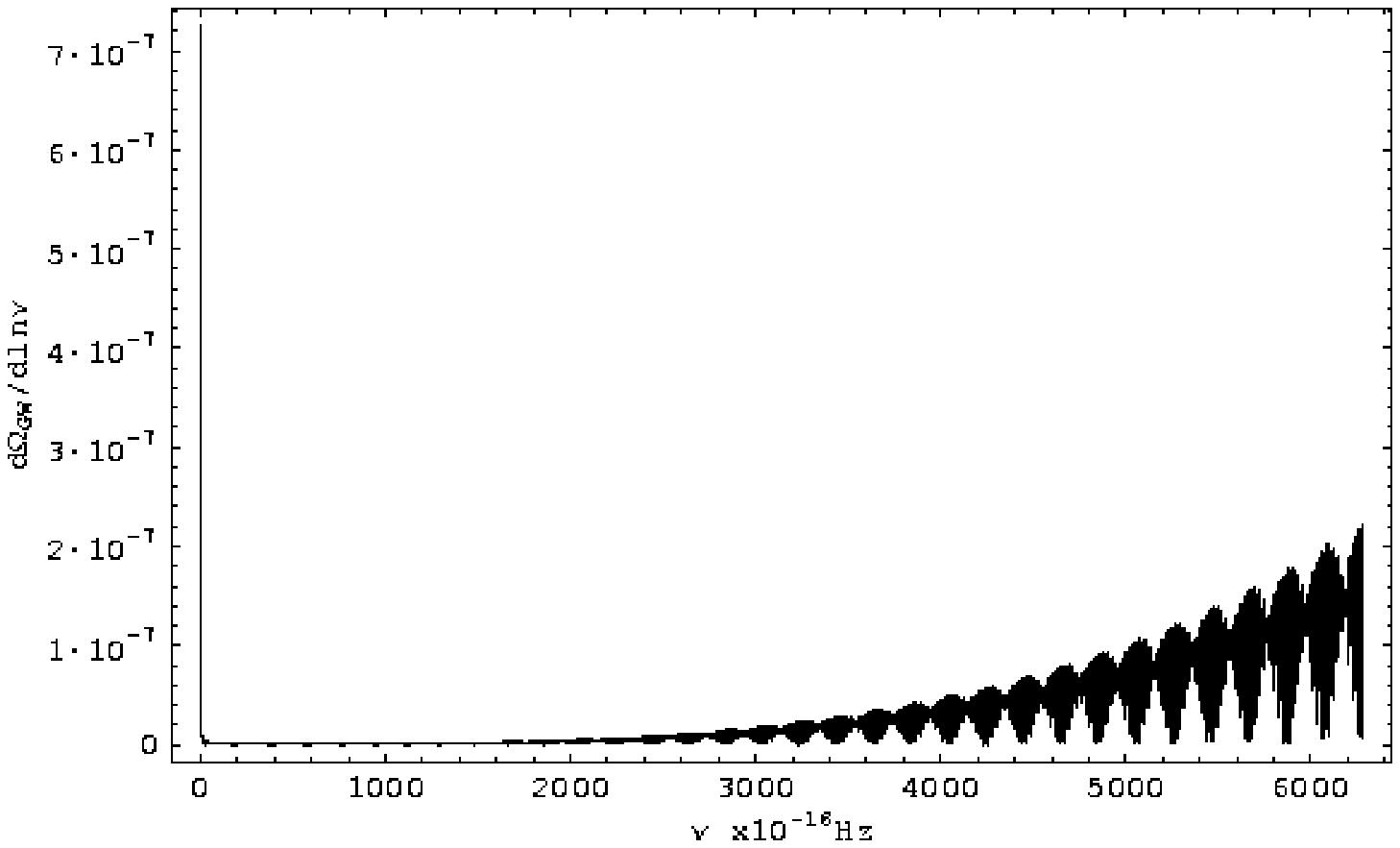}
    {\footnotesize {\bf Figure 10.} GW spectrum for
       $\omega=-0.5$,
       $\Omega_{x_0}=0.7$ and
       $\Omega_{m_0}=0.3$.}
  \end{minipage} \hfill &
  \begin{minipage}[b]{0.47\linewidth}
    \includegraphics[width=\linewidth]{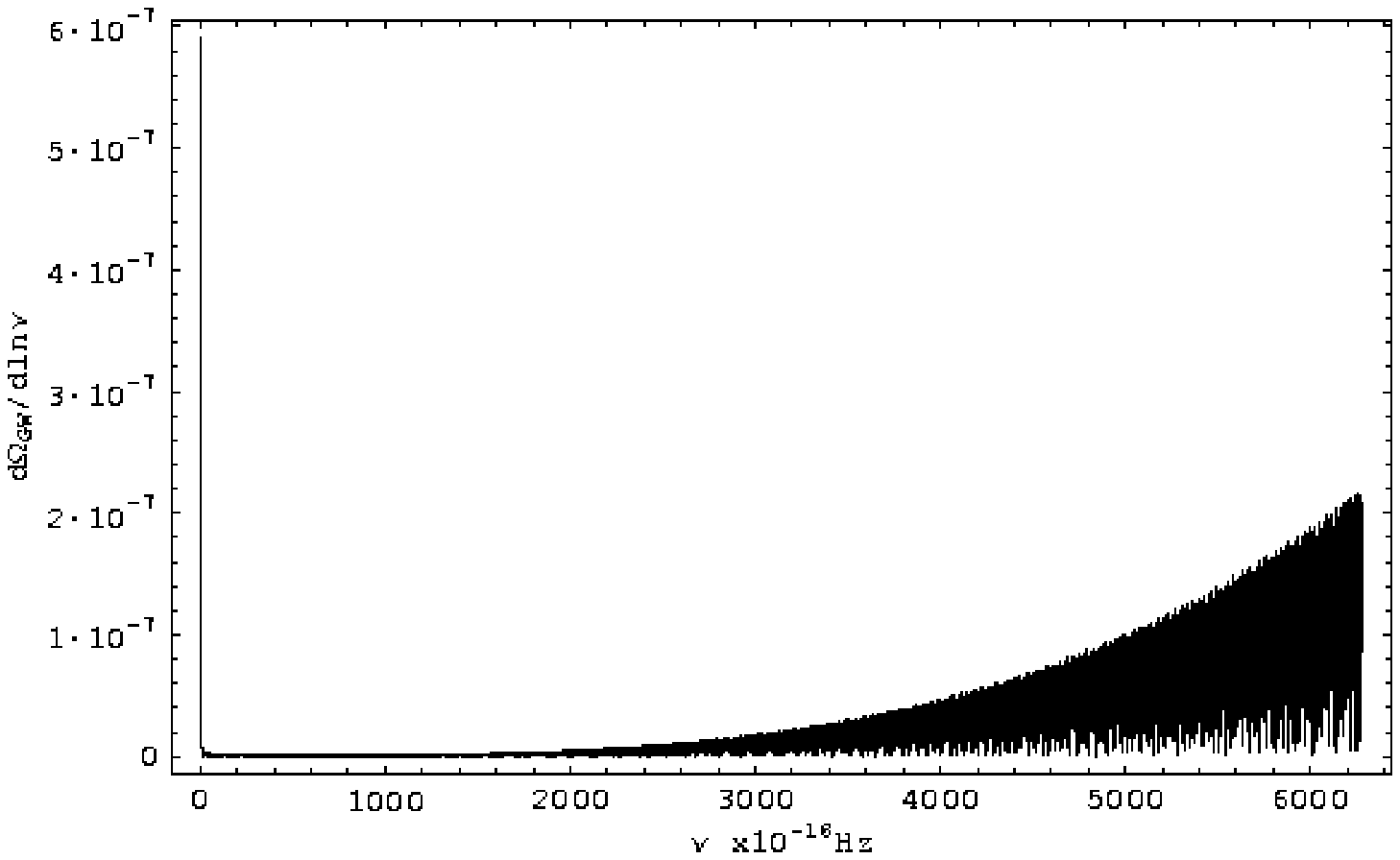}
    {\footnotesize {\bf Figure 11.} GW spectrum for
       $\omega=-10$,
       $\Omega_{x_0}=0.7$ and
       $\Omega_{m_0}=0.3$.}
  \end{minipage} \hfill
\end{tabular} \\\vspace{5mm}

(iv) The X-fluid is represented in plots (10) and (11). The later
is the case of phantom energy, with $\omega=-10$, while the former
is an intermediate situation between the pressureless and the
$\Lambda$ cases. We notice that they also {\it do not} differ from
each other.

\section{Conclusions}
In this work we have compared three important dark energy models
in the context of the gravitational wave spectra that they are
able to induce and we observe that the models are strongly
degenerated in the range of frequencies studied -- from $10^{-18}$
up to $10^{-15} Hz$. This degeneracy is in part expected, since
the models must converge to each other when some particular
combinations of parameters are considered. In addition, for the
Chaplygin gas, the negligible dependence on the density parameter
$\Omega_{c_0}$ is consistent with the idea of a unified dark
component \citep{Fabris:2004,Gorini:2004}. Since the gas
interpolates both dark energy and matter, the suppression of the
dark matter density parameter (see, plots (4-9)) should not affect
the spectrum, as observed. However, it is interesting how it seems
to be insensitive to  $\alpha$ as well. This result indicates that
this model is more sensitive to $\bar{A}$ than to $\alpha$. The
X-fluid component contribution seems to be completely degenerated
and must be better studied in the near future, specially at larger
frequency ranges.

Unfortunately, no direct observational data from cosmological
gravitational waves is available up to now, but still it is
important to investigate the relation between these gravitational
waves produced or amplified by the presence of the dark energy
component and the CMB temperature and polarization anisotropies
which are already observationally determined \citep{Bennett:2003}.
This correlation between CMB and gravitational waves will be
properly studied once we have a larger range of frequencies in the
GW spectra.


\section{Acknowledgments}
M.S.S. and E.M.G.D.P. acknowledge financial support from the
Brazilian Agencies FAPESP and CNPq.

\bibliography{bib}

\IfFileExists{\jobname.bbl}{}
 {\typeout{}
  \typeout{******************************************}
  \typeout{** Please run "bibtex \jobname" to optain}
  \typeout{** the bibliography and then re-run LaTeX}
  \typeout{** twice to fix the references!}
  \typeout{******************************************}
  \typeout{}
 }

\end{document}